\documentclass[article, prl, preprint, nobibnotes, secnumarabic, amssymb, superscriptaddress, aps,longbibliography]{revtex4-2}
\setcitestyle{super} 

\usepackage{graphicx}
\usepackage{amsmath}
\usepackage{upgreek}
\usepackage{color}
\usepackage{setspace}
\usepackage{hyperref}
\hypersetup{colorlinks,breaklinks,
            urlcolor=[rgb]{0,0,0.64},
            linkcolor=[rgb]{0,0,0.64},
            citecolor=[rgb]{0,0,0.64}}
\usepackage{nccmath}
\usepackage{amssymb}
\usepackage{lineno}

\begin{document}

\title{Room-temperature non-volatile optical manipulation of polar order in a charge density wave}

\author
{Qiaomei Liu,$^{1\ast}$ Dong Wu,$^{2\ast\dagger}$ Tianyi Wu,$^{1}$ Shanshan Han,$^{2}$ Yiran Peng,$^{3}$ Zhihong Yuan,$^{4}$ Yihan Cheng,$^{1}$  Bohan Li,$^{2}$ Tianchen Hu,$^{1}$ Li Yue,$^{1}$ Shuxiang Xu,$^{1}$ Ruoxuan Ding,$^{1}$ Ming Lu,$^{2}$ Rongsheng Li,$^{1}$ Sijie Zhang,$^{1}$ Baiqing Lv,$^{5}$ Alfred Zong,$^{6}$ Yifan Su,$^{7}$  Nuh Gedik,$^{7}$ Zhiping Yin,$^{3}$ Tao Dong,$^{1}$ Nanlin Wang $^{1,2\dagger}$\\
\vspace{12pt}
\normalsize{$^{1}$International Center for Quantum Materials, School of Physics, }\\
\normalsize{Peking University, Beijing 100871, China}\\
\normalsize{$^{2}$Beijing Academy of Quantum Information Sciences, Beijing 100913, China}\\
\normalsize{$^{3}$Department of Physics and Center for Advanced Quantum Studies,}\\
\normalsize{Beijing Normal University, Beijing 100875, China}\\
\normalsize{$^{4}$School of Physics and Information Engineering, Shanxi Normal University, Taiyuan 030031, China}\\
\normalsize{$^{5}$Tsung-Dao Lee Institute, Shanghai Jiao Tong University, Shanghai 200240, China}\\
\normalsize{$^{6}$Department of Chemistry, University of California, Berkeley, California 94720, USA}\\
\normalsize{$^{7}$Department of physics, Massachusetts Institute of Technology, }\\
\normalsize{Cambridge, Massachusetts 02139, USA}
\\
\vspace{12pt}
\normalsize{$^\ast$These authors contributed equally to this paper}
\\
\normalsize{$^\dagger$To whom correspondence should be addressed; E-mail: wudong@baqis.ac.cn; nlwang@pku.edu.cn}
\\
}
\date{\today}

\maketitle

\noindent\textbf{Utilizing ultrafast light–matter interaction to manipulate electronic states of quantum materials is an emerging area of research in condensed matter physics. It has significant implications for the development of future ultrafast electronic devices. However, the ability to induce long-lasting metastable electronic states in a fully reversible manner is a long-standing challenge. Here, by using ultrafast laser excitations, we demonstrate the capability to manipulate the electronic polar states in the charge-density-wave material EuTe$_{4}$ in a non-volatile manner. The process is completely reversible and is achieved at room temperature with an all-optical approach. Each induced non-volatile state brings about modifications to the electrical resistance and second harmonic generation intensity. The results point to layer-specific phase inversion dynamics by which photoexcitation mediates the stacking polar order of the system. Our findings extend the scope of non-volatile all-optical control of electronic states to ambient conditions, and highlight a distinct role of layer-dependent phase manipulation in quasi-two-dimensional systems with inherent sublayer stacking orders.}

\section{Introduction}
Materials that exhibit long-range electronic order lie at the heart of condensed matter physics and usually display emergent states such as multiferroics, charge density waves (CDW), and high-temperature superconductivity. Photoinduced metastable states are frequently  observed in these systems, but they are usually short-lived and revert to their initial states within a few picoseconds to microseconds through a combination of electronic, thermal, and lattice relaxation processes \cite{DeLaTorre2021,Bao2022,Dong2022,LSCO2011,floquet,WT2019,STOThz,Kogar2020,In2020,WTao2021,YTO2023}. When the photoinduced states persist permanently, i.e., in a non-volatile way, the phenomena have attracted particular interests and are considered technologically significant. Recent advances have revealed a few cases of ultrafast,  photoinduced non-volatile electronic order transitions at low temperatures in quantum materials, where the underlying lattice remains intact \cite{Stojchevska2014,Zhang2016,Mn-prl19}, which have deeply expanded our understanding of nonequilibrium dynamics. However, despite these advances, achieving non-volatile optical manipulation of electronic states under ambient conditions remains a long sought-after issue, particularly in an all-optical, reversible manner.

Here we report the discovery of non-volatile, all-optical manipulation of electronic states at room temperature in EuTe$_{4}$, a rare polar CDW semiconductor characterized by an inherent sublayer stacking order. We focus on femtosecond (fs) above-gap excitation and monitor electric polarization through second harmonic generation (SHG) experiments. The findings reveal a remarkable nonthermal process for photoinduced non-volatile switching of the polarization state, involving the reversible generation and annihilation of SHG signals, accompanied by resistance changes purely through optical methods. Our results indicate a critical role of layer-specific photocontrol in phase transformations, particularly relating to changes in sublayer stacking order.

EuTe$_{4}$ is a quasi-two-dimensional semiconducting CDW material \cite{ET2019,ET2022}. Its crystal structure consists of planar Te-sheets separated by insulating EuTe slabs (Fig.\,1a). At room temperature, the CDW distortion of the square Te-sheets yields a near 1 $\textit{a}$ $\times$  3 $\textit{b}$ $\times$  2 $\textit{c}$ superlattice \cite {ET2022,ET3}, where the superlattice along the $\textit{b}$-axis is incommensurate. In a single Te-sheet, the CDW develops with a modulation vector $\textbf{q}$ directed along the $\textit{b}$-axis, which is perpendicular to its polarization vector $\textbf{e}$ (the direction of Te atomic displacement, along the $\textit{a}$-axis; Fig.\,1b). These combined atomic displacements lead to Te-bond rearrangements and the formation of regular in-plane Te-trimers. Within the single Te-sheet, the in-plane inversion symmetry is broken due to the CDW distortion, with the polar axis oriented along the $\textit{a}$-axis (see Fig.\,1b, right panel). In other words, the formation of Te-trimers collectively generates a net polarization along the $\textit{a}$-axis in a single CDW Te-sheet. Such an improper polarization \cite{impro2008} is a concomitant property of CDW order in EuTe$_{4}$. As illustrated in Fig.\,1c, the parallel stacking of polarization in adjacent Te-sheets results in the breaking of spatial inversion symmetry along the $\textit{a}$-axis and hence a polar order in the EuTe$_{4}$ CDW state.

In the experiments reported here, EuTe$_{4}$ is excited with 800 nm laser pulses from a Ti: Sapphire laser amplifier system, with a pulse duration of 35 fs and a repetition rate of 1 kHz (see Methods). A writing beam is set to induce phase transitions (see Fig.\,1d). Structural symmetry information is monitored by continuously rotating the linear polarizer in the incoming 800 nm beam and measuring the reflected 400 nm second harmonic projected onto the \textit{a-b} plane along the [001] crystallographic direction with a second polarizer. The electrical resistance is measured using four-electrode method. We select a near-normal-incidence reflection geometry, and thus the SHG light probes in-plane symmetry properties (see Supplementary Note 1). The measured angle-dependent SHG patterns match well with the XRD-measured noncentrosymmetric polar point group ($\textit{C$_{2v}$}$) of the CDW EuTe$_{4}$ (see Supplementary Figure 1). The static temperature-dependent SHG intensity of the pristine sample exhibits a typical thermal hysteresis behavior (Fig.\,1e), following the same trend as the ones found in resistance and CDW superlattice peak intensity \cite{ET2019,ET2022}.

\section{Results}
Optical experiments are performed as a function of writing pulse fluence on nanometer-thick EuTe$_{4}$ samples (see Methods). As shown in Fig.\,2 and 3, a writing pulse is able to either suppress or enhance the SHG intensity, with simultaneous switching of the electrical resistance. The phenomena are found to be prominent already at room temperature and remain detectable up to 350 K (see Supplementary Figure 2). The photoinduced phases are non-volatile, persisting indefinitely (verified for up to 3 weeks) after the writing pulse is turned off.

We observe two distinct excitation regimes separated by a threshold fluence F$_{cR}$ $\sim$ 6.8\,mJ/cm$^{2}$ of writing pulse, denoted as Regime-W for weak excitations and Regime-S for strong excitations (see Fig.\, 2 and 3, and more details for F$_{cR}$ in Supplementary Figure 3). In Regime-W, starting from the initial state at 300 K on cooling branch, a single writing pulse results in a non-volatile state characterized by a increase in both resistance and SHG intensity (see Fig.\,2a and b). The maximum values of these quantities are achieved at a writing intensity of approximately 1.5\,mJ/cm$^{2}$ (see Fig.\,2e and f, red curves). Above this intensity, a single writing pulse induces states where both resistance and SHG intensity gradually decrease, culminating in the complete suppression of SHG intensity at a fluence of approximately 4\,mJ/cm$^{2}$ (see Fig.\,2e, red curve). Conversely, when starting from 300 K on heating branch, a single writing pulse in Regime-W decreases both the resistance and SHG intensity (see Fig.\,2c and d). We further measure the shot-by-shot resistance and SHG intensity with sequentially increasing the writing pulse fluence. Fig.\,2e and f show the results measured at 300 K from different thermal branches. Empirically, in Regime-W, the complete removal of photoinduced states can be effectively achieved through thermal annealing, returning the system to its original ground state (see more details in Supplementary Figure 4).

Once the writing fluence exceeds F$_{cR}$ (Regime-S), a single writing pulse creates a new non-volatile state with no SHG signal but with resistance increase by several orders of magnitude (see Fig.\,3). Subsequently, by applying the same writing beam at a lower intensity level (in a range of 4-5 mJ/cm$^{2}$), a train of thousands of pulses induces a new polar state characterized by low resistance and a unique SHG pattern. Analysis of this SHG pattern reveals a strongly enhanced nonlinear susceptibility tensor component $\chi_{aaa}$ (see Supplementary Figure 5 and Table 1). This photoinduced state likely results from a process in which photoexcitations above F$_{cR}$ induce a strong light–matter interaction, leading to additional atomic displacements in the Te-sheets that amplify the nonlinearity along the $\textit{a}$-axis in an irreversible manner. Further investigations using structure-sensitive techniques are required to clarify this mechanism. In Regime-S, an all-optical rewritable phase transition is discovered: when a single pulse with fluence exceeding F$_{cR}$ is applied, the system transitions into a state with a vanishing SHG signal and high resistance; subsequent lower-flunence pulses recover the system to a polar state with strong SHG signal and low resistance (see Fig.\,3b and c). Additional details for repeatability of the phase transitions in different samples can be found in Supplementary Figure 6. We note that there must be a significant energy barrier separating different photoinduced non-volatile states, which may be held in the domain walls or within the polarization states, providing strong stabilization force against thermal excitation. The energy barriers can be observed in both Regimes-W and Regimes-S by examining the fluence thresholds required for triggering the states switching (see Fig. 2e and Supplementary Figure 7). Importantly, the in-plane CDW superlattice remains unchanged with a same period across all non-volatile states, as confirmed by transmission electron microscopy (TEM) experiments (see Supplementary Note 3 and Figure 8). 

To gain insights into the microscopic dynamics of the above phase transitions in EuTe$_{4}$ system, we investigate the photoinduced nonequilibrium state using time-resolved SHG spectroscopy. Fig.\,4a shows the ultrafast measurements of SHG intensity at 300 K on pristine heating branch, before inducing the non-volatile phases. Following the pump pulse excitation, the SHG intensity exhibits a sharp drop within $\sim$ 0.2 picoseconds (ps), followed by a three-step relaxation process: a first positive increase till $\sim$ 1 ps, before the typical lattice thermalization timescale, a subsequent negative relaxation within time constant $\tau_{2}$  $\sim$ 1 ps, and finally, a slow lattice relaxation that gradually slows down, approaching divergence until non-volatile phase transition (see Fig.\,4b). Here the sharp drop of SHG intensity within $\sim$ 0.2 ps can be ascribed to a transient valence-band depopulation, where pump excitation depletes electrons from the states responsible for the strong dipole signal \cite {GaAs1}. The unusual hump process centered at $\sim$ 1 ps deviates from the common two-temperature relaxation model \cite{ferro,MoTe2019}, suggesting complex microscopic interactions in EuTe$_{4}$ system. Following this hump process, the relaxation rate becomes highly dependent on fluence, indicating a transition towards non-volatile states as the system approaches the phase transition point (see Fig. 4b). 

Time-resolved measurements of the SHG intensity $I_{SH} \propto [\epsilon_{0}\chi^{(2)}E^{2}(t)]^{2}$ reveal the magnitude of time-dependent second-order susceptibility \cite{shg2,MoS13,ferro}. Symmetry analysis indicates that only two independent third-rank tensor components, $\chi_{aaa}$ and $\chi_{abb}$, contribute to the in-plane SHG signal (see Supplementary Note 2). To further explore the anomalous dynamics, we measure the pump-induced relative change of SHG intensity contributed by $\chi_{aaa}$ and $\chi_{abb}$ tensor components separately. As shown in Fig.\,4c and 4d, upon excitation, each spectrum displays the characteristic hump feature centered at 1 ps during the relaxation process. Interestingly, the SHG spectra corresponding to $\chi_{aaa}$ show a significant transient enhancement, with the peak nearly doubling the initial magnitude under 1.1  mJ/cm$^{2}$ excitation. In contrast, the hump peaks in the $\chi_{abb}$  spectra are typically smaller than the initial values (see Fig.\,4d). The transient SHG pattern under 1.1  mJ/cm$^{2}$ excitation at 1 ps is illustrated in Fig.\,4e (see more detail in Supplementary Figure 9). This significant anisotropic response with pronounced increase along the $\textit{a}$-axis, occurs shortly after the electron decay process \cite {ET2024} (see Supplementary Figure 10). This implies that the system's electric polarization is highly influenced by nonthermal process associated with photoexcited electrons, pointing to strong anharmonic couplings among the electrons, CDW and polar order\cite {pin15,ferro}. In this context, it is important to note that the pulse-duration-dependent measurements (see Methods) support the nonthermal nature of the observed photoinduced non-volatile phenomena. As shown in Fig.\,4f, at the same writing fluence, the non-volatile change of SHG intensity scales inversely with the writing pulse duration. The result suggests that the nonthermal anharmonic process plays a crucial role in non-volatile polarization switching.

Thus, several notable features of the photoinduced non-volatile states in EuTe$_{4}$ can be concluded: (1) Photoexcitations above specific thresholds can induce permanent changes of SHG intensity and resistance; (2) Giant resistance and new SHG pattern are available in Regime-S; (3) Time-resolved SHG spectra reveal strong nonlinear anharmonic effects sensitive to photoexcitations; and (4) The in-plane CDW modulation remains intact.

\section{Discussion}

The photoinduced permanent change in SHG intensity reflects the change of the material's polarity. There are three potential scenarios that could explain this polarity change in EuTe$_{4}$: (1) a phase transition between the CDW polar state and the high-temperature non-CDW nonpolar state; (2) a phase transition within different polarity states while maintaining the CDW order; and (3) polar domain formation with the CDW order preserved. In this third scenario, the SHG intensity might decrease due to the destructive interference between second harmonic beams reflected from opposite polarization domains, and vice versa \cite{ferro}. The first mechanism can be ruled out as the CDW modulation consistently remains in the non-volatile phases. As discussed further, our observations are intrinsically linked to the dynamics of photoinduced layer-specific polarization inversion, covering both the second and third scenarios.



In view of layered structure of EuTe$_{4}$, there are six inequivalent Te-sheets stacking alternately along the $\textit{c}$-axis in a CDW supercell, including two mono-layers and two bi-layers (see Supplementary Figure 12). Each Te-sheet exhibits electric polarization along the $\textit{a}$-axis, either positive or negative in principle,  depending on the direction of the CDW distortion. The overall polarity in EuTe$_{4}$ is thus determined by the relative stacking order of these Te-sublayers,  potentially resulting in various spatial symmetries and multiple polar phases \cite{MoS13,stack2017,stack2102,stack2301,stack2023}(see Supplementary Figure 13). For instance, in addition to the polar phase, an anti-polar phase with inversion symmetry is also reasonable based on our crystallographic modeling, which uses the same XRD data obtained from pristine samples (see Supplementary Table 2).

Upon photoexcitation, the relevant electronic states accessible by 1.55 eV photons are primarily derived from Te-p orbitals, which are split into subbands due to the CDW formation. These subbands mainly originate  from the Te mono-layers and bi-layers, forming nested manifolds of conduction and valence bands in an energy scale of about $\pm$ 1 eV near the Fermi level \cite{ET2019,ET2022,ET2024}. Photoexcitation of these Te-p orbitals initially creates equal numbers of electrons and holes in conduction and valence bands, leading to a nonequilibrium state in each Te-sheet by rapid intraband thermalization via carrier scattering. The electronic dynamics of this nonequilibrium state have been investigated through time- and angle-resolved photoemission measurements \cite{ET2024}. The results reveal that Te mono-layers and bi-layers exhibit distinct CDW orders, with the bi-layers displaying a stronger CDW intensity due to additional interactions within the bi-layers. Moreover, moderate photoexcitation firstly collapses the CDW energy gap in Te mono-layers, indicating that distinct fluences are required to disrupt the CDW orders in Te mono-layers and bi-layers.

Once the writing fluence exceeds the threshold required to melt the CDW condensate, ultrafast phase inversion dynamics are universally observed in CDW systems, corresponding that the CDW distortion overshoots to the opposite direction  \cite {Tb2010,KMO14,Pinv1,WTao2021,pinver2}. A similar phase inversion phenomenon has also been identified in ferroelectric materials \cite {ferro}. Drawing on these studies and building on our understanding of the polar CDW state in EuTe$_{4}$, we speculate that layer-specific phase inversion dynamics occur in EuTe$_{4}$ system. When a moderate photoexcitation above a certain threshold is applied, the relatively weak CDW and polar order in the Te mono-layers will collapse first, thereafter leading to a polarization overshoot to the opposite direction. Unlike other materials, where phase inversion is typically dissipative, the overshoot phase gets permanently trapped in EuTe$_{4}$. If the non-volatile polarization inversion occurs only in partial Te sublayers, it will induce a structural phase transition to a new state with polarization configuration distinct from the initial state. As such, with the six inequivalent Te-sheets stacking in one CDW super-cell, it could theoretically result in multitudes of structural phases through polarization inversion in different Te-sheets (see Supplementary Figure 13).

For a simple illustration, we construct a bipartite polarization system with an asymmetric potential energy landscape. Considering only the interlayer coupling between adjacent CDW Te mono-layers and bi-layers in EuTe$_{4}$, a two-level potential energy surface is formed, as described in the literature \cite {ET2019}. When the anharmonic polarization coupling is included, there will be a lopsided modification to the potential energy landscape \cite {pin15,ferro,Te2022}. Fig.\,4g depicts such a conceptual anharmonic two-level potential energy surface. For a ground state A with parallel polarization configuration, an excitation with intensity exceeding a threshold can induce a transition into metastable state B through typical layer phase inversion. Conversely, for the system in state B, a relative weak excitation over the shallow potential barrier can drive a transition back to state A, which has a stronger electric polarization. This sketch of reaction coordinate effectively captures the experimental trends observed in both Regime-W and S, where higher writing fluences reduce the SHG intensity while weaker fluences enhance it (see Fig.\,2 and 3).

In the actual EuTe$_{4}$ system, a variety of structural phases with diverse polarization configurations can be statistically generated through the layer-specific polarization inversion, and the interlayer couplings are more complex \cite {ET2024}. As well, the system exhibits notable nonlinear anharmonicity, which is highly sensitive to photoexcitation (see Fig. 4\,c-e). While it is not feasible to accurately model the intricate order parameters that require further study, one can still propose the existence of a highly corrugated and anharmonic potential energy landscape, containing multiple metastable states. In this scenario, the system's overall behavior mirrors that of the two-level system mentioned above: high fluences tend to excite more states with weaker averaged electric polarization, while low fluences lead the excited states back into polar phases with stronger electric polarization by overcoming shallow barriers.

It is worth noting that photoinduced domains would naturally form through phase transitions following percolation principles \cite {Mn-prl19,Zhang2016}. Domain growth is also a well-established phenomenon accompanying photoinduced phase inversion dynamics in other 2D CDW systems, where macroscopic domain walls typically form parallel to the surface at the boundary between the inverted and original phase \cite{Tb2010,Pinv1,pinver2,WTao2021}. Considering the dynamics of photoinduced phase inversion in EuTe$_{4}$, forming microscopically identical domains require simultaneous polarization reversal of all Te-sheets, which is a rare situation. Instead, for phase inversion induced in only partial Te-sheets, the phase transition and the domain with different phase would be concomitantly favored, a process strongly influenced by the spatial variation in writing fluence levels along the penetration path.


In summary, we demonstrate that above-band-gap femtosecond excitation enables permanent polarization phase switching in an intrinsically stacking-ordered material at room temperature. Our research reveals a nonthermal dynamics of photoinduced layer-specific phase inversion, favoring transitions among various stacking-ordered phases. The interlayer coupling and strong anharmonicity are expected to play an important role in the observed phenomena. While the exact microscopic mechanism still needs to be clarified, our non-volatile switching dynamics clearly differ from other methods of manipulating polarization phases, such as ultrafast heating \cite {LNO2022,LNB05}, bulk photovoltaic effect \cite {BFO18}, optically-induced strain \cite {STO2019} or strong excitation near epsilon-near-zero regime \cite {ENZ}. Our findings highlight the necessity for further research utilizing more advanced spatio-temporal techniques, such as femtosecond single-shot spectral analyses, to elucidate the underlying mechanisms behind this photoinduced switching phenomena in this typical layered system.

Considering the merits of nonvolatility, operation at ambient conditions, all-optical manipulation, electronic ordering and reversible switching, our finding in EuTe$_{4}$ holds potential for device applications. The finding is a rare example and will further motivate the search for new generations of room-temperature, ultrafast and non-volatile memory elements in electronically ordered materials.

\section{Methods}

\textbf{Sample growth and device fabrication}

EuTe$_{4}$ single crystals are grown via solid-state reaction using Te as a flux \cite{ET2019}. High-purity Eu lumps (99.999 \%) and Te granules (99.999 \%) are mixed in a stoichiometric ratio of about 1 : 15 and sealed in an evacuated fused silica tube. The mixture is heated at 850 $^\circ$C for 2 days in a muffle furnace, then slowly cooled to 415 $^\circ$C over 100 hours. It is held at 415 $^\circ$C for one week before decanted using a centrifuge. The crystals are planar-shaped with dark, mirror-like surfaces. EuTe$_{4}$ flakes, typically in thickness less than the penetration depth $\sim$22 nm, are mechanically exfoliated from bulk crystals and transferred onto sapphire substrates. The four electrodes are fabricated through standard ultraviolet lithography and electron beam evaporation techniques.\\

\textbf{Experimental Method}

All ultrafast laser beams are derived from a Ti: Sapphire Amplifier system (Spitfire Ace). The laser pulse has a central wavelength of 800 nm, a pulse duration of 35 fs, and a repetition rate of 1 kHz. The writing beam is focused on the sample with a spot diameter of about 700 $\mu$m, ensuring complete coverage of the sample. The resistance of EuTe$_{4}$ flake is recorded in situ through standard four-electrode method using a Keithley 2400 Source-Meter. The static and time-resolved SHG measurements are operated in a near-normal optical reflection geometry. The incident probe light (800 nm) is firstly tuned to circularly polarization using a quarter-wave plate, with further polarization control achieved by rotating a linear polarizer. The reflected second harmonic (400 nm) signal is filtered through short-pass and bandpass filter groups and collected using a photomultiplier tube photodetector (PMT, PMM01, Thorlabs), with the bias voltage set at 0.65 V. The PMT output is sent to a lock-in amplifier with modulation of 373 Hz via a mechanical chopper.\\
The pulse duration is measured using a commercial scanning auto-correlator (GECO, Light Conversion). It is adjusted by varying the relative distance of the grating pair inside the Amplifier laser system, which mainly affects the second-order dispersion. Besides, the output laser pulse keeps a standard Gaussian shape with constant power. \\
\vspace{12pt}
\\
\noindent\textbf{Data Availability Statement}: Source data are provided with this paper. Other datasets generated and/or analyzed during the current study are available from the corresponding author upon request.\\

\bibliography{bib} .\\

\noindent\textbf{Acknowledgements}:
We thank YC Ma, AA Liu for fruitful discussions. This work is supported by National Natural Science Foundation of China (Grant Number 12488201 by N.L.W., 12274033 by D.W., 12374063 and 23Z990202580 by B.Q.L.), National Key Research and Development Program of China (Grant number 2022YFA1403901 by N.L.W.), Shanghai Natural Science Fund for Original Exploration Program (Grant Number 23ZR1479900 by B.Q.L.), US Department of Energy (Basic Energy Sciences, Materials Sciences and Engineering Division-data analysis), the Gordon and Betty Moore Foundation's EPiQS Initiative grant GBMF9459 (instrumentation) received by N.G.\\

\noindent\textbf{Author Contributions Statement}: N.L.W., D.W. and Q.M.L. conceived the study. Q.M.L., D.W., S.S.H., B.H.L., T.C.H., L.Y., S.X.X., R.S.L. and S.J.Z. synthesized the single crystals, performed the optical and electrical transport measurements. T.Y.W., S.S.H., Q.M.L., D.W., M.L., Y.R.P., Z.H.Y., Y.H.C., R.X.D., B.Q.L., A.Z., Y.F.S. and N.G. performed theoretical analyses and calculations. D.W., Q.M.L.and N.L.W. analyzed the data and wrote the manuscript with critical inputs from all other authors. \\

\noindent\textbf{Competing Interests Statement}: The authors declare no competing interests. \\

\clearpage
\newpage

\clearpage
\newpage

\begin{figure}[t]
  \centering
  \includegraphics[width=15cm]{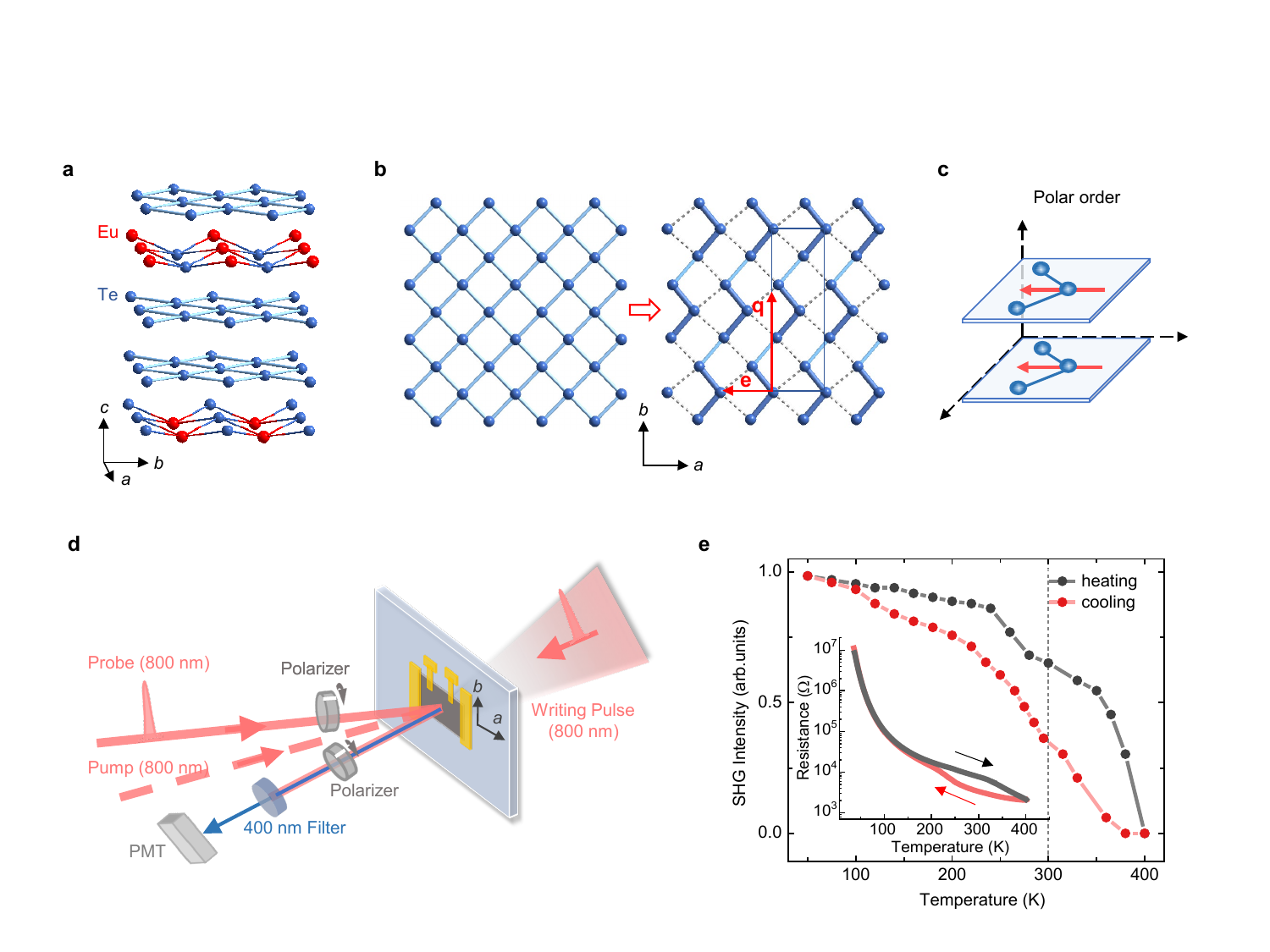}\\
  \caption{\textbf{Schematic description of photo-manipulation of the polar electronic state in EuTe$_{4}$}. \textbf{a.} The averaged crystal structure of EuTe$_{4}$, consisting of layered square Te-sheets. \textbf{b.} The Peierls instability in square Te-sheet leads to the formation of distorted Te-trimers. The CDW vector $\textbf{q}$ is oriented along the $\textit{b}$-axis, which is perpendicular to the polarization vector $\textbf{e}$. The right panel illustrates the in-plane inversion symmetry breaking in a single Te-sheet. \textbf{c.} The schematic of stacking sequence of the polarized Te-sheets for the formation of polar order, where Te-trimers in adjacent Te-sheets align in the same direction. \textbf{d.} A Sketch of the experimental setup, where PMT is the photomultiplier tube. \textbf{e.} Temperature dependence of SHG intensity of the pristine sample. It shows a thermal hysteresis between the heating and cooling processes, which is coincident closely to the hysteresis observed in the resistance curve in the inset. The SHG intensities are measured in the parallel channel with an incident angle of $\sim 58^{o}$.} 
  \label{Fig1}
\end{figure}

\begin{figure}[t]
  \centering
  \includegraphics[width=15cm]{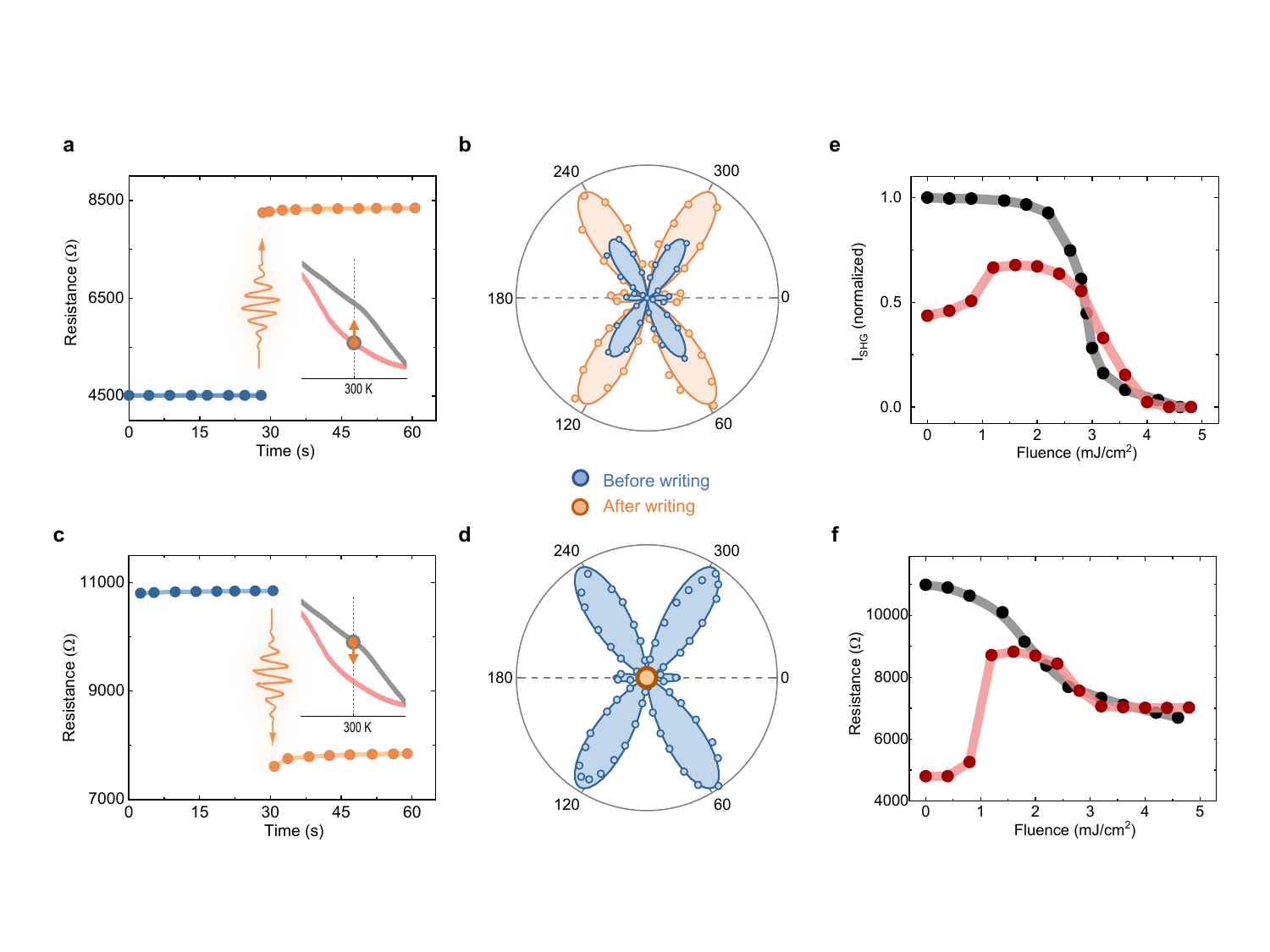}\\
  \caption{\textbf{Non-volatile switching of SHG intensity and  resistance induced by a 35-fs writing pulse at 800 nm in weak excitation regime}. \textbf{a.}, \textbf{b.} Starting from the state at 300 K on cooling branch (see the inset in (\textbf{a})), a single weak pulse, e.g., 1.5 mJ/cm$^{2}$, enhances both resistance (\textbf{a}) and SHG intensity (\textbf{b}). \textbf{c.}, \textbf{d.} Conversely, at 300 K on heating branch, a more intense pulse, e.g., 4.5 mJ/cm$^{2}$, suppresses both resistance (\textbf{c}) and SHG intensity (\textbf{d}). \textbf{e.} Shot-to-shot SHG intensity and \textbf{f.} resistance as a function of writing fluence. The data points are measured by increasing fluence in sequence starting from the cooling (red curves) or heating branch (black curves) respectively.  All measurements are operated at 300 K. The SHG patterns of (\textbf{b}) and (\textbf{d}) are measured in parallel channel.}
  \label{Fig2}
\end{figure}


\begin{figure}[t]
  \centering
  \includegraphics[width=15cm]{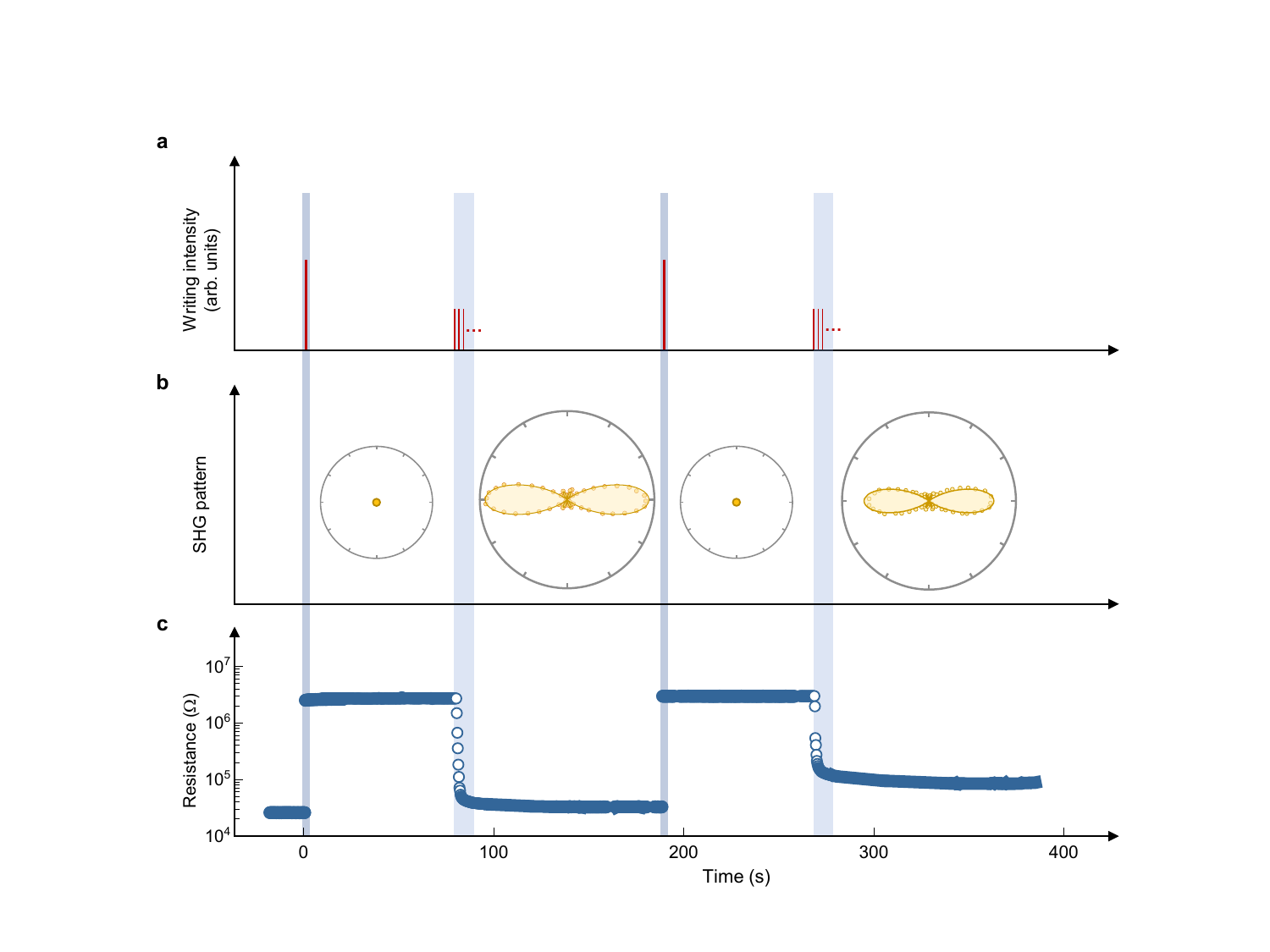}\\
  \caption{\textbf{Non-volatile switching of SHG intensity and resistance induced in strong excitation regime}. A high intensity pulse above F$_{cR}$ drives the EuTe$_{4}$ system into a new non-volatile phase. \textbf{a.} Depiction of the writing pulse number and intensity used in sequence: a single shot with strong excitation intensity of about 7.5 mJ/cm$^{2}$, and the subsequent ten thousands shots (10 s, 1 kHz) with moderate intensity of $\sim$ 4.5 mJ/cm$^{2}$. \textbf{b.} The measuered SHG pattern and intensity switching after exposure to the writing pulses corresponding to (\textbf{a}). A new SHG pattern is induced under the moderate excitations. The SHG signals are measured in parallel channel. \textbf{c.} The corresponding sharp resistance switching.}
  \label{Fig3}
\end{figure}

\begin{figure}[t]
  \centering
  \includegraphics[width=15cm]{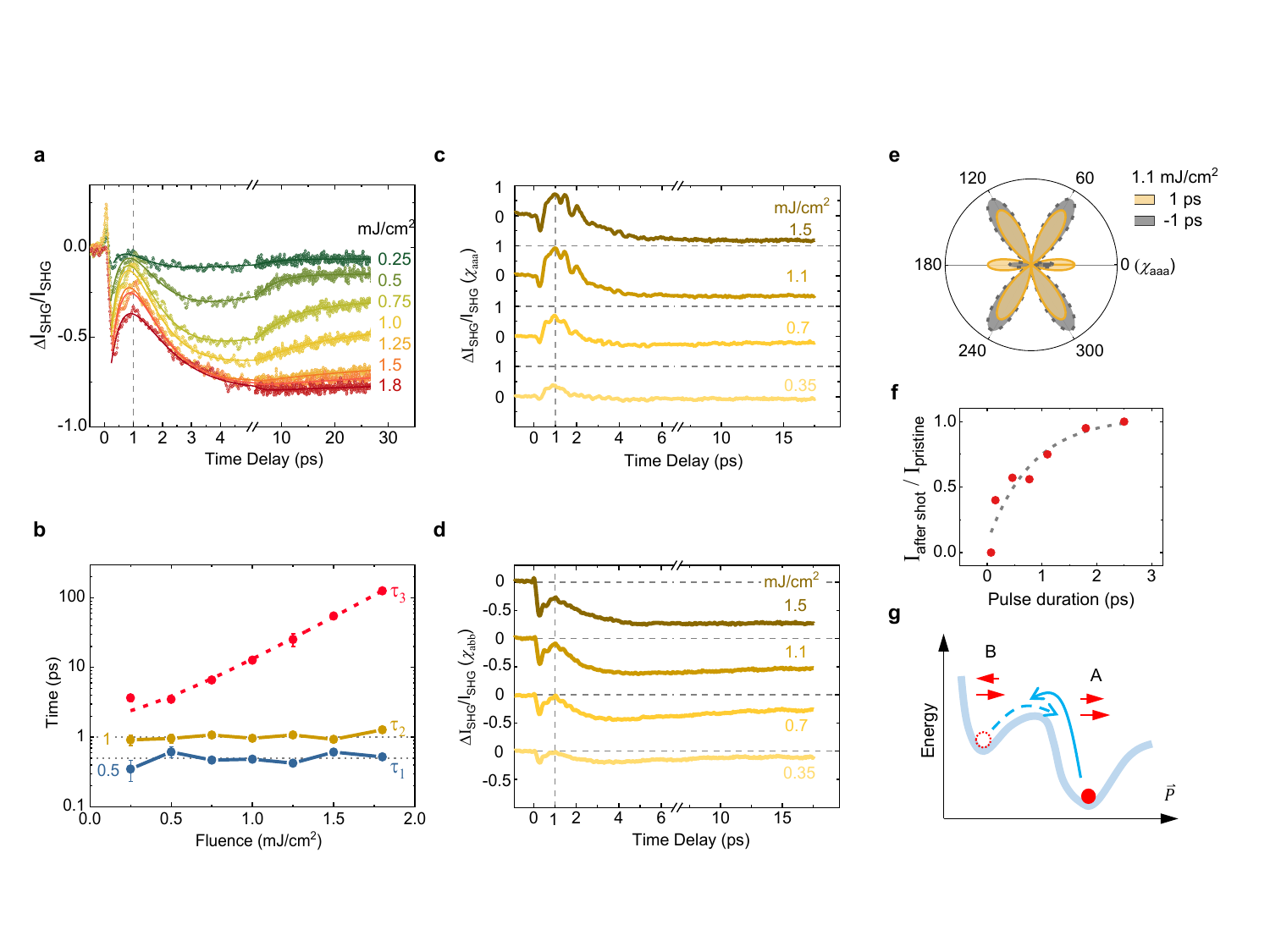}\\
  \caption{\textbf{Dynamical properties of the nonequilibrium states in EuTe$_{4}$}. \textbf{a.} Time traces of pump-induced relative change of SHG intensity at various pump fluences before non-volatile switching. After photoexcitation, a three-step relaxation process is observed and the solid lines represent fitted curves using a three-exponential function. The SHG spectra are measured at 300 K on pristine heating branch and at an incident angle of 58° in parallel channel. \textbf{b.} The three time constants from exponential fits as a function of pump fluence extracted from the time traces in (\textbf{a}). The vertical error bars correspond to standard error derived from the fits in (\textbf{a}). \textbf{c.}, \textbf{d.} The relative change of SHG intensity solely contributed by $\chi_{aaa}$ (\textbf{c}) and $\chi_{abb}$ (\textbf{d}) tensor component.  \textbf{e.} The transient SHG pattern measured at 1 ps (pale yellow) after photoexcitation in parallel channel, compared to the pristine one before time zero (-1 ps, light gray). \textbf{f.} The photoinduced non-volatile change of SHG intensity as a function of writing pulse duration. Each data point is measured at different pristine sample areas and the non-volatile states are induced by exposure to writing pulses (10 shots, 3 mJ/cm²). For pulse duration exceeding approximately 2 ps, the SHG intensity remains nearly unaffected. The dash curve is a guide to the eye. \textbf{g.} Schematic of a bipartite polarization system with asymmetric two-level potential energy landscape, where a bidirectional phase transition occurs via photoinduced layer-specific polarization inversion. The short red arrow represents weak polarization, while the long arrow indicates strong polarization.}


  \label{Fig4}
\end{figure}

\clearpage





\end{document}